\documentclass[preprint,proceedings]{rmaa}

\SetYear{2016}
\SetConfTitle{XV Latin American Regional IAU Meeting LARIM}

\title{Astroparticle Techniques: Colombia Active Volcano Candidates for Muon Telescope Observation Sites}

\author{
H. Asorey\altaffilmark{1,2}, A. Balaguera-Rojas\altaffilmark{2}, L.A. N\'u\~nez\altaffilmark{2,3}, J.D. Sanabria-G\'omez\altaffilmark{2}, C. Sarmiento-Cano\altaffilmark{2}, M. S\'uarez-Dur\'an\altaffilmark{2}, M. Valencia-Otero\altaffilmark{2} and A. Vesga-Ram\'irez\altaffilmark{2}.
}

\altaffiltext{1}{Laboratorio Detecci\'on de Part\'iculas y Radiaci\'on, Centro At\'omico Bariloche \& Instituto Balseiro, Bariloche, Argentina.}

\altaffiltext{2}{Escuela de F\'isica, Universidad Industrial de Santander, Bucaramanga, Colombia.}

\altaffiltext{3}{Departamento de F\'isica, Universidad de Los Andes, M\'erida, Venezuela.}

\suppressfulladdresses

\listofauthors{H. Asorey}
\listofauthors{A. Balaguera-Rojas}
\listofauthors{L. A. N\'{u}\~{n}ez}
\listofauthors{J.D. Sanabria-G\'omez}
\listofauthors{C. Sarmiento-Cano}
\listofauthors{M. S\'uarez-Dur\'an}
\listofauthors{M. Valencia-Otero}
\listofauthors{A. Vesga-Ram\'irez}
\listofauthors{Asorey, H.}
\listofauthors{Balaguera-Rojas, A.}
\listofauthors{N\'{u}\~{n}ez, L. A. }
\listofauthors{Sanabria-G\'omez, J.D. }
\listofauthors{Sarmiento-Cano, C. }
\listofauthors{S\'uarez-Dur\'an, M.}
\listofauthors{Valencia-Otero, M. }
\listofauthors{Vesga-Ram\'irez, A. }
\addkeyword{Astroparticle physics}
\addkeyword{High Energy Astrophysics}
\addkeyword{Particle detectors}
\addkeyword{Volcano muography}

\begin{document}
% Typeset article header
\maketitle 

\boldabstract{We discuss a methodology to identify observation points for muongraphy of active Colombian Volcanoes and it is found that only Cerro Mach\'in can be studied.}

The MUTE (Muon Telescope) is a project to design, construct and operate a hybrid detector (2 scintillator panels + a water Cherenkov detector) to measure integrated flow of muons crossing a rock mass ($\approx 100$\, m) in Colombian active volcanoes.

Muongraphy measures the attenuation of cosmic muon flux making it possible to build images of volcanic inner structure at the top of the edifice. It uses the same basic principles of standard medical radiography\citep{MarteauEtal2012,Tanaka2014}.

As the particles measured at ground are produced by the interaction of cosmic rays with atmospheric elements, its modulation needs to be carefully corrected by considering those factors that could affect the integrated flux at Earth surface --mainly geomagnetic conditions, atmospheric reaction and detector response--\citep{AsoreyEtal2015B}. From this simulation chain we estimate the background muon flux, as a function of its arrival direction and energy, incident on geological structures. 

\begin{figure}[h!]
 \includegraphics[width=\columnwidth]{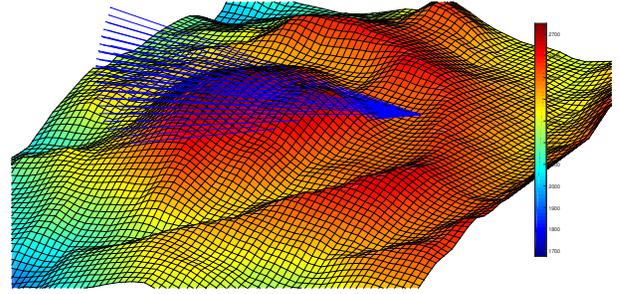}
 \caption{Cerro Mach\'in with ones of its raytracing}
 \label{Figura1}
\end{figure}

Muons cross matter and lose energy described by
$-\frac{dE}{d\varrho}=a(E) + b(E)E$, where $a(E)$ and $b(E)$ are functions of the material and $\varrho(L)$ the density integrated along trajectories. If we chose homogeneous standard rock, (i.e $<Z/A>=0.5$ and density $2.65$\,g\,cm$^{-3}$), the distances traveled by muons and its integrated flux density, --depending on their energy and arrival directions-- are determined, considering volcano topographies and its surroundings, (See Figure \ref{Figura1}). 
\begin{figure}[h!]
 \includegraphics[width=\columnwidth]{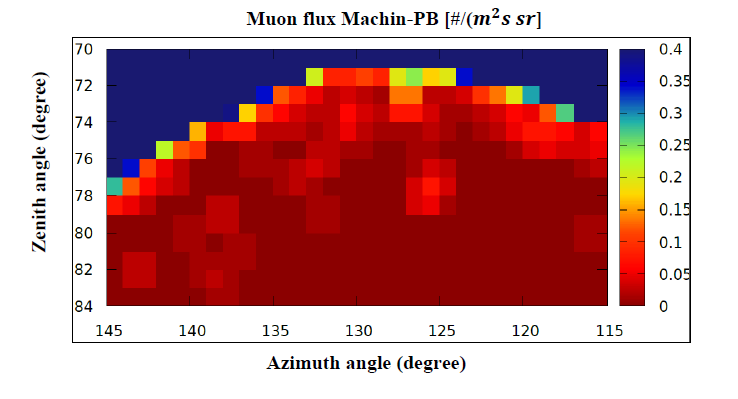}
	\caption{Muon flux at 4$^{\circ}$29'31.14"N, 75$^{\circ}$22'48.31"W, Cerro Mach\'in.}
 \label{fig:simple}
\end{figure}

After having analyzed four Colombian volcanoes --Nevado del Ruiz, Galeras, Cerro Mach\'in and Cerro Negro-Chiles-- we have determined that only Cerro Mach\'in is feasible to be studied by muon tomography from two points: 4$^{\circ}$29'39.53"N, 75$^{\circ}$23'17.04"W, and 4$^{\circ}$29'31.14"N, 75$^{\circ}$22'48.31"W. 

This project has been partially funded by ColCiencias, under contract FP44842-051-2015 and also supported by VIE Universidad Industrial de Santander.

\end{document}